\documentclass[prb]{revtex4-1}

\usepackage{graphicx}

\usepackage[utf8]{inputenc}

\usepackage{amsfonts}
\usepackage{amssymb}

\newcommand{\be}{\begin{equation}}
\newcommand{\ee}{\end{equation}}
\newcommand{\bea}{\begin{eqnarray}}
\newcommand{\eea}{\end{eqnarray}}
\newcommand{\Dd}{\mathrm{d}}
\renewcommand{\doi}[1]{DOI:~\href{http://doi.org/#1}{#1}}
\usepackage{siunitx}

\begin{document}

\title[Shapiro time delay and equivalence principle]{The Shapiro time delay and the equivalence principle}

\author{M P{\"o}ssel}
\address{Haus der Astronomie and Max Planck Institute for Astronomy,\\ K{\"o}nigstuhl 17, 69124 Heidelberg, Germany}
\email{poessel@hda-hd.de}

\begin{abstract}
The gravitational time delay of light, also called the Shapiro time delay, is one of the four classical tests of Einstein's theory of general relativity. This article derives the Newtonian version of the Shapiro time delay from Einstein's principle of equivalence and the Newtonian description of gravity, in a manner that is accessible to undergraduate students and advanced high-school students. The derivation can be used as a pedagogical tool, similar to the way that simplified derivations of the gravitational deflection of light are used in teaching about general relativity without making use of the more advanced mathematical concepts. Next, we compare different general-relativistic derivations of the Shapiro time delay from the Schwarzschild metric, which leads to an instructive example for the challenges of formulating the post-Newtonian limit of Einstein's theory. The article also describes simple applications of the time delay formula to observations within our solar system, as well as to binary pulsars.
\end{abstract}
\pacs{45.20.dh}
\maketitle 

\section{Introduction}

The fourth classical test of general relativity is variously known as the {\em gravitational time delay} or the {\em Shapiro delay}, after Irwin I. Shapiro, who predicted the effect in 1964\cite{Shapiro1964} and conducted the first direct observations a few years later, using radar echoes within our solar system.\cite{Shapiro1968,Shapiro1971} The Shapiro delay has been measured with great accuracy using space probes with onboard transponders. One such measurement, using the Cassini probe in orbit around Saturn, yields the most stringent current test confirming that the propagation of light (both light deflection and the gravitational time delay) follows the predictions of Einstein's theory of gravity.\cite{Will2014} More recently, the Shapiro effect has played an important role in tests of general relativity in an environment with strong gravitational fields, namely near White Dwarfs or neutron stars, making use of binary pulsars.

Several derivations of the effect at the undergraduate level exist, which are based on the Schwarzschild metric in either Schwarzschild or in isotropic coordinates.\cite{Bruckman1993} I am not aware of attempts to derive the gravitational time delay without recourse to the metric, at the same level as numerous existing derivations for the deflection of light.\cite{Treder1981,Koltun1982,Will1988,Lotze2005} The aim of this article is to fill the gap, presenting a simple derivation, accessible to undergraduates and advanced high-school students, which deduces a location-dependent speed of light from the equivalence principle, and pays particular attention to the notion of the post-Newtonian and Newtonian limits of general relativity.  The result of the simplified derivation deviates from the full general-relativistic result by a factor 2; this is the same as for Newtonian calculations of the gravitational deflection of light.\cite{Lerner1997} Section \ref{Sec:Ambiguities} extends the analysis to descriptions using the Schwarzschild metric, comparing different text-book derivations of the Shapiro effect and pointing out ambiguities that can be used to prepare students for the necessity of defining a rigorous framework for post-Newtonian physics. Section \ref{Sec:Data} provides the information that will enable students to understand astronomical applications of the gravitational time delay, both in the solar system and for binary pulsars.

\section{Equivalence principle and coordinate speed of light}
\label{Sec:EquivalenceLight}

General relativity contains both special relativity and Newtonian gravity as limiting cases: Take away all masses and all gravitational waves, so that all terms containing the gravitational constant $G$ can be neglected, and you obtain special relativity. Consider only speeds that are much slower than the speed of light, $v\ll c$, and you obtain Newton's theory of gravity (and the associated laws of classical mechanics). The challenge of describing general-relativistic effects in a weak-gravity situation is the need to combine elements from Newtonian gravity and special relativity in a suitable way, at least approximately. This is made more difficult by general relativity's underlying covariance: Coordinates do not automatically carry physical meaning. 

A simple --- albeit incomplete --- way is to combine gravity as described by Newton's theory with light propagation as derived from the Einstein Equivalence Principle (abbreviated EEP in the following). The EEP says that, in a free-falling reference frame, locally, the laws of physics are the same as those of special relativity. ``Locally'' means that the region is small enough, both spatially and in terms of the length of time our description is meant to cover, for us to be able to neglect the influence of tidal forces: changes of gravity from location to location, or over time, are too small to significantly affect the accuracy of our description. The laws of physics to which the EEP is applicable are not taken to include the laws governing gravity (that is left to the Strong Equivalence Principle), but they do include the special-relativistic law that has light moving along straight lines at the constant speed $c_0$. In the following, we will always write $c_0$ for the special-relativistic value for the speed of light, reserving non-annotated $c$, in particular with an argument, such as $c(r)$, for the local coordinate speed of light in a specified coordinate system.

Specifically, consider a static situation, with test particles moving in a Newtonian gravitational field.
We choose Cartesian spatial coordinates $\vec{x}$ in which the geometry of space is Euclidean and the space coordinates faithfully reproduce physical lengths, and we choose a time coordinate $t$ such that, in these coordinates, the Newtonian equations for gravity hold at least for test particles with $v\ll c$. We will call that combination of time coordinate and space coordinates our {\em Newtonian coordinates}.

In that setting, a test particle with trajectory $\vec{x}(t)$ will experience the gravitational acceleration
\be
\frac{\Dd^2 \vec{x}}{\Dd t^2} = -\vec{\nabla}\Phi(\vec{x}(t)),
\label{Eq:GravAcceleration}
\ee
where $\Phi$ is the gravitational potential. For simplicity, we will also assume spherical symmetry, with a potential $\Phi(r)$ depending only on the usual radial coordinate $r=|\vec{x}|\equiv\sqrt{x^2+y^2+z^2}$ for some chosen origin, although all of the following local arguments are also valid when considering a direction perpendicular to the local equipotential surface. We will also assume that $\lim_{r\to\infty}\Phi(r)=0,$ so that far from the center, physics is described in good approximation by the laws of special relativity. In particular, light in such regions will move at the usual constant speed $c_0$. 

The simplest case that complies with all our assumptions is that of a point mass with mass $M$ located at the origin $\vec{x}=\vec{0}$, which corresponds to a gravitational potential
\be
\Phi(r) = -\frac{GM}{|\vec{x}|} = -\frac{GM}{r}.
\label{Eq:PointMassPotential}
\ee

Given our assumptions, there are two sets of properties that are not yet fixed: Consider a set $\cal S$ of clocks, each at rest at a fixed location in the Newtonian coordinate system. We will refer to observers who are at rest relative to one of those clocks, who make use of the clock's proper time for their time measurements and measure spatial distances using the local Newtonian space coordinates, as {\em stationary observers}. The proper time rates for the $\cal S$ clocks are not determined by our Newtonian framework, since purely Newtonian physics only knows about coordinate time, taken to represent absolute time. Also, within Newtonian physics alone, limited as it is to $v\ll c$, we cannot make any deductions about the propagation of light.

We will now derive the proper times of clocks at rest in our gravitational field, and the laws of light propagation in the shape of a location-dependent coordinate speed of light $c(\vec{x})$ in the Newtonian coordinates, with the following strategy: Whenever we want to calculate effects of gravity in the regime $v\ll c$, we will use the Newtonian formulas. Whenever we want to study the propagation of light, we will introduce suitable local inertial frames in free fall, and use the EEP to describe how light propagates in those frames. We will go back and forth between the two kinds of description --- Newtonian coordinates vs. the coordinates of the local inertial frames --- as appropriate, linking their results in order to derive the local proper time rate and the local coordinate speed of light at each location.

\begin{figure}[htbp]
\begin{center}
\includegraphics[width=0.5\textwidth]{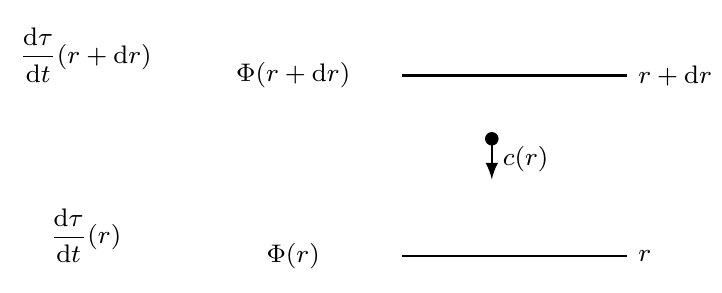}
\caption{Motion of light from one location to another}
\label{Fig:GravFigure}
\end{center}
\end{figure}
To this end, consider two locations in our scenario, one at radial coordinate value $r$, the other infinitesimally higher at $r+\Dd r$, as shown in Fig.~\ref{Fig:GravFigure}. At each location, we have written down the value of the potential $\Phi$ and the rate
\be
\frac{\Dd\tau}{\Dd t}(r)
\ee
relating the Newtonian time coordinate $t$ to the proper time $\tau$ of the $\cal S$ clock at rest at fixed $r$. In order to see that the coordinate speed of light, which we have denoted $c(r)$, only depends on $r$, consider an inertial system $\cal I$ in free fall that, at the event $\cal E$ when the light passes a specific clock, is at rest in the Newtonian coordinate system. At $\cal E$, length measurements in $\cal I$ coincide with length measurements by an observer attached to one of the $\cal S$ clocks, since both frames are at relative rest, and the length of an infinitesimal time interval as measured in $\cal I$ is the same as the proper time $\tau(r)$ of the $\cal S$ clock. In the free-fall inertial frame, light moves at the speed $c_0$; thus, as measured by the $\cal S$ clock and the local space coordinates, the observer at rest relative to the $\cal S$ clock, using the spatial coordinates and its own proper time, will find the light to be moving at the same speed,
\be
 \left|\frac{\Dd \vec{x}}{\Dd\tau}\right| = c_0.
\ee
With our assumption of spherical symmetry, the rates of our stationary $\cal S$ clocks can only depend on $r$. In consequence, the coordinate speed of light in our Newtonian coordinates must be
\be
c(r)=\left|\frac{\Dd \vec{x}}{\Dd t}\right| = \left|\frac{\Dd \vec{x}}{\Dd\tau}\right|\cdot\frac{\Dd\tau}{\Dd t}(r)= c_0\cdot\frac{\Dd\tau}{\Dd t}(r)
\label{Eq:LightSpeedR}
\ee
Now assume that a light signal is travelling from an observer at rest at $r+\Dd r$ to an observer at rest at $r$. If the light wave leaves the upper observer at the time $t$, it will arrive at the lower observer at the time 
\be
t + \Dd t=t + \frac{\Dd r}{c(r)}.
\label{Eq:AccTime}
\ee 
Introduce two inertial frames in free fall: ${\cal I}_1$ is at rest relative to the upper stationary observer at $r+\Dd r$, at the exact event when the light is emitted, and ${\cal I}_2$ is at rest relative to the lower stationary observer at $r$ at the event where the light is received. By this definition, an observer in ${\cal I}_1$ will measure the same wavelength $\lambda_E$ for the light as the upper stationary observer at the time of emission, and ${\cal I}_2$ the same wavelength $\lambda_R$ as the lower stationary observer at the time the light is received.

Between the time of emission and the time of reception, the free-falling frame ${\cal I}_1$ has gained a speed $\Dd v$ in our Newtonian r-t coordinate system. According to (\ref{Eq:GravAcceleration}), applied to purely radial motion, and inserting the time interval (\ref{Eq:AccTime}), we have
\be
\Dd v = \frac{\Dd^2 r}{\Dd t^2}\cdot \Dd t = -\frac{\Dd\Phi}{\Dd r}\Dd t = -\frac{\Dd\Phi}{\Dd r}\cdot  \frac{\Dd r}{c(r)} = -\frac{\Dd\Phi}{c(r)}.
\ee
This is only the coordinate speed, however. The observer in ${\cal I}_2$, whose time is momentarily the same as the proper time $\tau$ of the stationary  observer at position $r$, has a measure of time that differs from the Newtonian coordinate time $t$, and will instead measure a small velocity $\Dd v_I$ for the inertial system ${\cal I}_1$ that is related to $\Dd v$ by
\be
\Dd v = \Dd v_I\cdot\frac{\Dd\tau}{\Dd t}=\Dd v_I\cdot \frac{c(r)}{c_0}.
\ee
By the EEP, we are allowed to use the laws of special relativity to express the relation of a wavelength for light as measured in ${\cal I}_1$ and the same wavelength as measured in ${\cal I}_2$: the two are linked by the Doppler formula, and since $\Dd v_I$ is infinitesimally small, we are allowed to apply the classical Doppler formula, which yields
\be
z \equiv \frac{\lambda_R-\lambda_E}{\lambda_E} = \frac{\Dd v_I}{c_0} = -\frac{\Dd\Phi}{c(r)^2}. 
\label{Eq:ZFormula}
\ee
This expression is related to the time it takes light to move past a clock: In ${\cal I}_1$, one wavelength of light will take the time $\Delta\tau =\lambda/c_0$ to move past at the speed $c_0$, and correspondingly for ${\cal I}_2$. But in our static Newtonian coordinates, one peak of the wave will take the same coordinate time to move from $r+\Dd r$ down to $r$ as the peak following it; in consequence, wave peaks that have been sent by the upper stationary observer a coordinate time interval $\Delta t$ apart will arrive at the location of the lower stationary observer the same coordinate time interval $\Delta t$ apart. Thus, the two proper time intervals in ${\cal I}_1$ and in ${\cal I}_2$ are proper time intervals on the local $\cal S$ clocks of the upper and the lower observer that correspond to the same coordinate time interval $\Delta t$, so that 
\be
\frac{\Dd\tau(r+\Dd r)}{\Dd \tau(r)} =\frac{\lambda_E}{\lambda_R} =  \frac{1}{1+z}\approx 1-z = 1+\frac{\Dd\Phi}{c(r)^2},
\label{Eq:RedshiftFormula}
\ee
where in the last step we have used eq.~(\ref{Eq:ZFormula}). With these results, and taking the presence of infinitesimal entities such as $\Dd r$ to imply the taking of appropriate limits, we can compute
\bea
\nonumber
\frac{\Dd c(r)}{\Dd r} &=& \frac{c(r+\Dd r)-c(r)}{\Dd r} \stackrel{(\ref{Eq:LightSpeedR})}{=} c_0\cdot\frac{\Dd\tau(r+\Dd r)-\Dd\tau(r)}{\Dd r\cdot\Dd t}\\[0.5em]
\nonumber
&=&\frac{\left[\frac{\Dd\tau(r+\Dd r)}{\Dd\tau(r)}-1
\right]}{\Dd r}\cdot c_0\cdot\frac{\Dd\tau(r)}{\Dd t} \stackrel{(\ref{Eq:LightSpeedR})}{=}\frac{\left[\frac{\Dd\tau(r+\Dd r)}{\Dd\tau(r)}-1
\right]}{\Dd r}\cdot c(r) \\[0.5em]
&\stackrel{(\ref{Eq:RedshiftFormula})}{=}& \frac{\left[\frac{\Dd\Phi}{c(r)^2}\right]}{\Dd r}\cdot c(r)=  \frac{1}{c(r)}\frac{\Dd\Phi}{\Dd r}.
\eea
Let us integrate up both sides from a specific $r$ value to infinity, remembering that
$\Phi(\infty)=0$ and $c(\infty)=c_0$. As a result, we obtain the equation
\be
c(r) = c_0\sqrt{
1+\frac{2\Phi(r)}{c_0^2}
}.
\label{Eq:LightSpeedFormula}
\ee
For a single spherically-symmetric mass, this comes out as
\be
c(r)= c_0\sqrt{
1-\frac{2GM}{rc_0^2}
} = c_0\sqrt{
1-\frac{R_{SSR}}{r}
},\label{Eq:LightSpeedFormulaSpherical}
\ee
where in the last equation, we have introduced the Schwarzschild radius corresponding to our mass $M$, defined as $R_{SSR}\equiv 2GM/c_0^2$. The Schwarzschild radius characterises the size of a spherical black hole of mass $M$, and more generally represents the typical length scale at which strong gravitational effects become important in a given situation.

Note that this result is in direct contradiction to what we would have expected, had we treated light like a Newtonian particle, governed by classical mechanics. The classical conservation of energy for light moving at the speed $c_0$ when infinitely far from a central mass would yield 
\be
c(r) = c_0\cdot \sqrt{1+\frac{2GM}{rc_0^2}},
\ee
which is larger than $c_0$ as the light approaches the mass. As we would expect from special relativity, Newtonian calculations give incorrect results when applied to entities travelling near or at the speed of light. This raises the question of how Newtonian calculations, such those of Soldner or Cavendish,\cite{Will1988} can yield results that are at least qualitatively (that is, up to an overall factor 2) correct. An important part of the answer is likely to be that, in order to reconstruct the shape of an orbit in the x-y plane, all that is needed is a valid expression for the ratio of the velocity components $v_x$ and $v_y$. Re-scaling both velocity components with the same location-dependent factor $c(r)$ does not change that ratio. In calculating the gravitational time delay, on the other hand, the correct form of $c(r)$ plays the central role.

\section{Newtonian gravitational time delay}
\label{Sec:NewtonianCalc}

\begin{figure}[htbp]
\begin{center}
\includegraphics[width=0.6\textwidth]{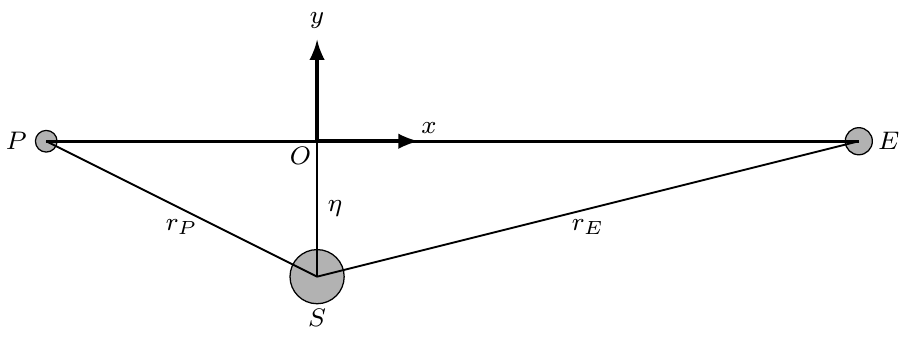}
\caption{Set-up for light from a planet $P$ passing the Sun $S$ on its way to Earth $E$}
\label{Fig:GTDSetup}
\end{center}
\end{figure}

With the result (\ref{Eq:LightSpeedFormulaSpherical}) for $c(r)$, let us calculate the gravitational time delay near a point mass, with Newtonian potential (\ref{Eq:PointMassPotential}). The simplest situation is that shown in Fig.~\ref{Fig:GTDSetup},
where we have light from a planet $P$ pass the Sun $S$ on its way to Earth $E$. We have approximated the trajectory of the light as a straight line from $P$ to $E$, and placed the origin of our x-y coordinate system at the point $O$ of the light's closest approach to the Sun. In these coordinates, the planet is at $(x_P, 0),$ the Earth at $(x_E,0)$ and the Sun at $(0,-\eta)$, where $\eta$ is called the {\em impact parameter}. The radius values are $r_E=\sqrt{x_E^2+\eta^2}$ and $r_P=\sqrt{x_P^2+\eta^2}$.

When light moving in the x direction at $y=0$ is at the coordinate position $x$, its distance from the Sun is $r=\sqrt{x^2+\eta^2},$ and from the expression (\ref{Eq:LightSpeedFormulaSpherical}) for $c(r)$, we know that the light will be moving at the coordinate speed
\be
\frac{\Dd x}{\Dd t} = c_0\left[
1-\frac{GM}{c_0^2\sqrt{x^2+\eta^2}}
\right],
\ee
where we have used the approximation $\sqrt{1+\xi}\approx 1+\xi/2$. In order to integrate this expression, we separate the variables $x$ and $t$ and, after having made use of the identity $(1-\xi)^{-1}\approx 1+\xi$, we obtain
\be
\Dd t = \frac{\Dd x}{c_0}\left[
1+\frac{GM}{c_0^2\sqrt{x^2+\eta^2}}
\right].
\label{Eq:DtDxc}
\ee
Integrating up from the time $t_P$ the light left the planet to the time $t_E$ the light reached Earth, we find that 
\be
t_E-t_P = \int\limits^{t_E}_{t_P} \Dd t = \int\limits^{x_E}_{x_P}\frac{\Dd x}{c_0} \left[1+\frac{GM}{c_0^2\sqrt{x^2+\eta^2}} \right] = \frac{x_E-x_P}{c_0} + \frac{GM}{c_0^3}\ln\left[
\frac{x_E+r_E}{x_P+r_P}
\right],
\label{Eq:TimeDelay1}
\ee
where we have made use of the integral formula 
\be
\int\frac{\Dd x}{\sqrt{x^2+a^2}} = \ln\left[
\sqrt{x^2+a^2} +x
\right]
\ee
which is straightforward to verify, albeit not so straightforward to derive from scratch. The linear term in (\ref{Eq:TimeDelay1}) is what we would have expected for light travelling at the constant speed $c_0$. The logarithmic term is the additional contribution of the gravitational time delay.

\begin{figure}[htbp]
\begin{center}
\includegraphics[width=0.6\textwidth]{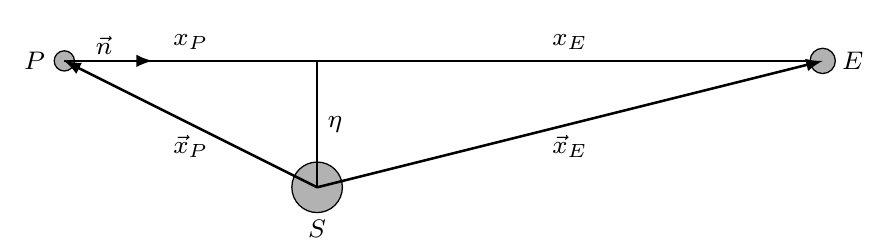}
\caption{Describing light propagation in terms of $\vec{x}_P$, $\vec{x}_E$ and the unit vector $\vec{n}$}\label{GTDSetup2}
\end{center}
\end{figure}

While the coordinate choice we made was particularly well-suited for the calculation, there is a way of re-writing our result in a somewhat different form, which is closer to the conventions in the literature,\cite{Will2014} using the quantities defined in Fig.~\ref{GTDSetup2}.
This time, we describe positions with the help of position vectors anchored at the Sun $S$. Introducing the unit vector $\vec{n}$ pointing from $P$ to $E$ along the direction of travel of the light, we have
$x_P =\vec{n}\cdot \vec{x}_P$ and $x_E =\vec{n}\cdot \vec{x}_E$. Multiply numerator and denominator in the argument of the logarithm in (\ref{Eq:TimeDelay1}) by $r_P-x_P$ and substitute the new position vectors, and the travel time becomes
\be
t_E-t_P= \frac{|\vec{x}_E - \vec{x}_P|}{c_0} + \frac{GM}{c_0^3}\ln\left[
\frac{(r_E +\vec{n}\cdot\vec{x}_E)(r_P -\vec{n}\cdot\vec{x}_P)}{\eta^2}\right].
\label{Eq:TimeDelay2}
\ee
Comparing this expression with a proper calculation in the framework of the post-Newtonian formalism for general relativity,\cite{Poisson2014} one finds that the Shapiro delay term is off by a factor 2: The general-relativistic effect is twice as large as the Newtonian effect calculated here. This is exactly the same as for simplified Newtonian derivations of the gravitational deflection of light; there, too, the general-relativistic deflection angle is twice as large as the angle predicted from a Newtonian perspective. We will follow up on that factor in section \ref{Sec:Ambiguities}.

Taken together, the results of this section suggest a possible way of teaching about the Shapiro delay in a setting where direct derivations from the Schwarzschild metric are not an option: Begin by presenting the simplified derivation developed in this section. This will yield a result that has the correct functional dependence on the geometry, but is off by an overall factor 2. Give the students the additional information that a more thorough derivation, which includes the curvature of space, will yield a result that includes the additional factor 2. After that statement, you can use the corrected formula, with the extra factor of 2, to consider applications such as the ones presented in section \ref{Sec:Data}, where the Shapiro time delay formula is used to compare predictions with data.

\section{Gravitational time delay, Schwarzschild metric and post-Newtonian ambiguities}
\label{Sec:Ambiguities}
Before we move on to applications, let us consider the origin of the factor 2 that comes out of more rigorous calculations. The results will also be of interest for teachers who derive the Shapiro time delay formula directly from the Schwarzschild metric, as it will highlight some of the challenges of teaching about post-Newtonian approximations in general.

\begin{figure}[htbp]
\begin{center}
\includegraphics[width=0.6\textwidth]{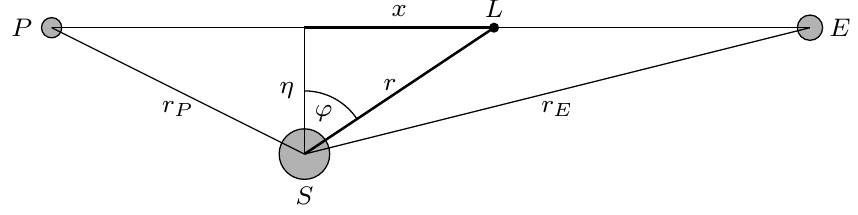}
\caption{Polar coordinates for light propagation near the Sun}
\label{Fig:PolarCoordinates}
\end{center}
\end{figure}

Consider a light-like geodesic in the Schwarzschild metric, written in Schwarzschild's original coordinates,
\be
\Dd s^2 = -\left[1-\frac{2GM}{c_0^2r}\right]c_0^2\Dd t^2 + \left[1-\frac{2GM}{c_0^2r}\right]^{-1}\Dd r^2
  +r^2(\Dd\theta^2 + \sin^2\theta\cdot\Dd\varphi^2).
  \label{Eq:SchwarzschildMetric}
\ee
In these coordinates, the (spherical) mass $M$ is located at the origin of the coordinate system. Let us choose the orientation of our angular coordinates so that the positions of the two planets involved, as well as the trajectory of the light, are all in the equatorial plane $\theta=\pi/2$. Fig.~\ref{Fig:PolarCoordinates} shows part of that plane, where the origin is now at the center of the Sun $S$.
Along the line $EP$, we have $x=r\sin\varphi$. Changing the $x$ coordinate on this line by $\Dd x$, we must change $r$ by $\Dd r=x/\sqrt{x^2+\eta^2}\cdot\Dd x$ and 
$\varphi$ by $\Dd\varphi=\Dd x[\eta/(\eta^2+x^2)].$ Inserting those two infinitesimal spatial coordinate changes into the metric (\ref{Eq:SchwarzschildMetric}) and setting $\Dd s^2=0$ as the condition for light-like geodesics, we obtain
\be
\Dd t =\frac{\Dd x}{c_0}\left[1+\frac{2GM}{c_0^2r} - \frac{GM}{c^2_0}\frac{\eta^2}{r^3}\right],
 \label{Eq:SchwarzschildDelay1}
\ee
where we have already made use of  $2GM/c_0^2\ll r$, which is valid in weak gravitational fields, to simplify the right-hand side. Compared with (\ref{Eq:DtDxc}), we find that there is indeed a new factor of 2 in the $1/r$ term. But there is also an extra term, which is proportional to $1/r^3$.

\begin{figure}[htbp]
\begin{center}
\includegraphics[width=0.45\textwidth]{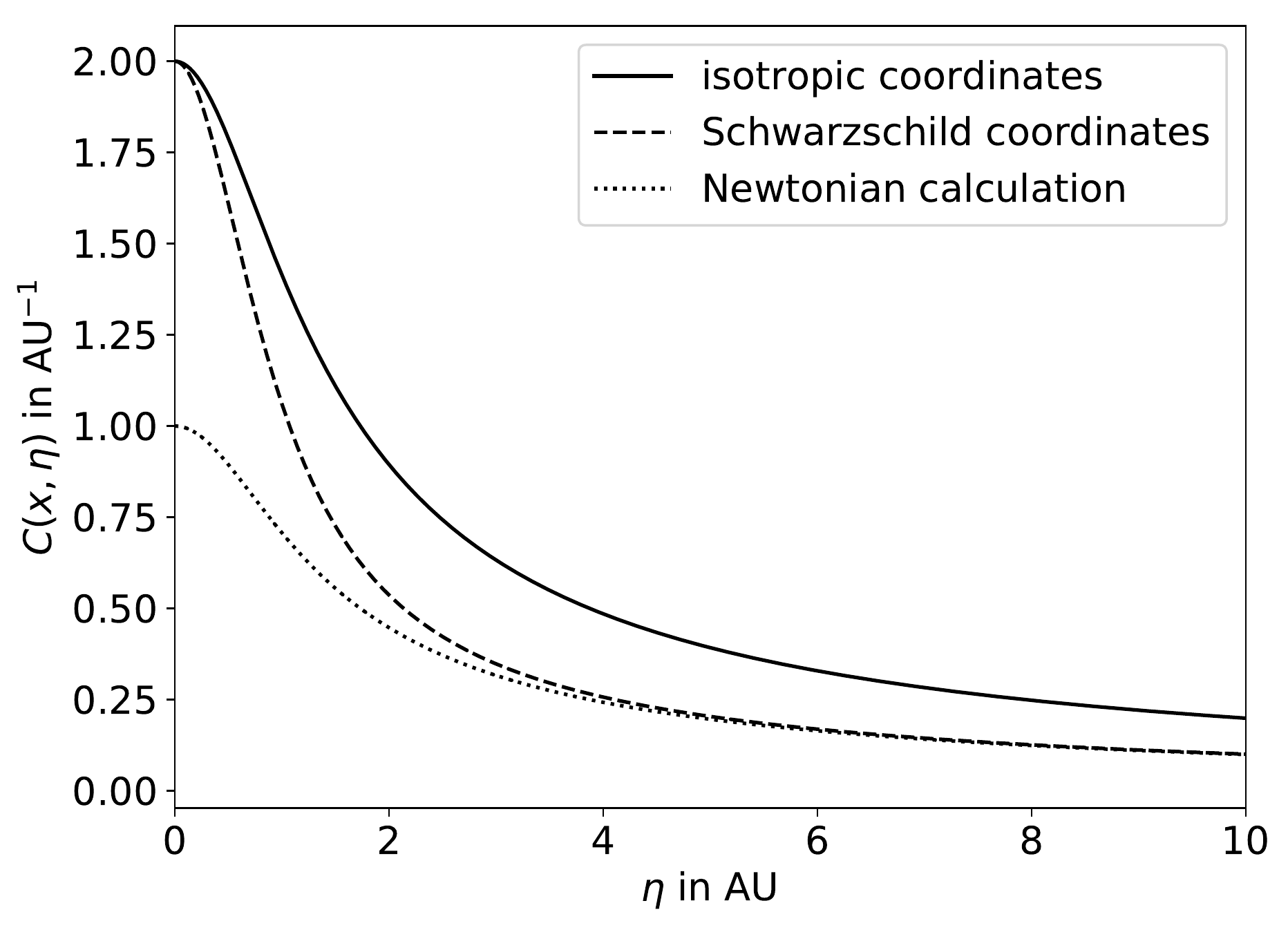}
\caption{The quantity $C(x,\eta)$ for $x=1\:\mbox{AU}$ and various values for $\eta$, comparing the Newtonian case, the result in Schwarzschild coordinates and the results in isotropic coordinates}
\label{Fig:SchwarzschildComparison}
\end{center}
\end{figure}

Before following up on this extra term, consider an alternative calculation. Define the new radial coordinate $\bar{r}$ via
\be
r=\left(
1+\frac{GM}{2c_0^2\bar{r}}
\right)^2\bar{r}.
\ee
Using this coordinate plus $t$ and the angular coordinates, we can write down the Schwarzschild metric in isotropic coordinates,
\be
\Dd s^2 = -\left[\frac{1-\frac{GM}{2c_0^2\bar{r}}}{1+\frac{GM}{2c_0^2\bar{r}}}\right]^2c_0^2\Dd t^2 + \left[1+\frac{GM}{2c_0^2\bar{r}}\right]^{4}\left[\Dd \bar{r}^2
  +\bar{r}^2(\Dd\theta^2 + \sin^2\theta\cdot\Dd\phi^2)\right].
\ee
Introducing Sun-centered Cartesian coordinates, with 
\be
\Dd\bar{r}^2 + \bar{r}^2(\Dd\theta^2 + \sin^2\theta\cdot\Dd\phi^2)=\Dd x^2
+\Dd y^2+\Dd z^2,
\ee
and again restricting ourselves to a situation where motion is in the x direction, so $\Dd y=\Dd z=0$,
we have
\be
\Dd t=\frac{\Dd x}{c_0}\frac{\left[1+\frac{GM}{2c_0^2\bar{r}}\right]^3}{\left[1-\frac{GM}{2c_0^2\bar{r}}\right]\:}
\approx \frac{\Dd x}{c_0}\left[1+\frac{2GM}{c_0^2\bar{r}}\right].
\label{Eq:SchwarzschildDelay2}
\ee
This differs from (\ref{Eq:DtDxc}) only by the extra factor 2 in the $1/\bar{r}$ term. There is no extra term proportional to $1/\bar{r}^3$.
Note that the differences between (\ref{Eq:SchwarzschildDelay1}) and  (\ref{Eq:SchwarzschildDelay2}) are considerable. Fig.~\ref{Fig:SchwarzschildComparison} shows plots of the quantity
\be
C(x,\eta) = \left[ c_0\frac{\Dd t}{\Dd x}-1\right]\left[
\frac{GM}{c_0^2}
\right]^{-1}
\ee 
at $x=1\:\mbox{AU}$ for various values of $\eta$, for all three of our cases: The Newtonian expression for $\Dd t/\Dd x$ that follows from (\ref{Eq:DtDxc}), the version that follows in Schwarzschild's original coordinates from (\ref{Eq:SchwarzschildDelay1}), and the version that follows from the calculation in isotropic coordinates, in (\ref{Eq:SchwarzschildDelay2}). As the figure demonstrates, for typical values within our solar system, the formulae give markedly different predictions.
The maximum value for $C(x,\eta)$ is the same for the calculation in Schwarzschild coordinates and that for isotropic coordinates, but as we move to larger $\eta$, the two results differ significantly. The Schwarzschild coordinate calculation can be seen to interpolate between the isotropic result for small $\eta$ and the Newtonian further out.

This discrepancy --- only one of the formulae can give the right approximation! --- is an instructive example of the ambiguities of taking the Newtonian limit. At infinity, both isotropic coordinates and Schwarzschild coordinates have the same flat-space limit. But the ways this limit is approached varies from one case to the other, leading to different correction terms of first order in $GM/(c_0^2L)$ with $L$ an appropriate length scale, such as $r$ or $\bar{r}$. 

While physical results are not coordinate-dependent, a Taylor expansion in specific coordinates, meant to capture general-relativistic corrections to Newtonian gravity, evidently {\em is} coordinate-dependent. As a result, even measurable, physical quantities like the Shapiro delay can come out differently, depending on the coordinate system chosen to calculate them. The comparison of the two different results  (\ref{Eq:SchwarzschildDelay1}) and  (\ref{Eq:SchwarzschildDelay2}) is a simple demonstration of the necessity of an approach to post-Newtonian corrections that is more systematic than merely collecting first-order terms in the given coordinate system --- a clear motivation for the development of a proper post-Newtonian expansion (whose details are, of course, too complicated to fit into an introductory undergraduate class).

Comparison with the post-Newtonian expansion\cite{Poisson2014} shows that the derivation based on isotropic coordinates is correct. It also gives an indication why, namely that the isotropic coordinates are similar to the harmonic coordinates that form the basis of the post-Newtonian formalism; in these harmonic coordinates, Einstein's equations can be re-formulated at least approximatively as a simple wave equation, which serves as the foundation of the series approximation at the heart of the post-Newtonian formalism.

General relativity text books tend to present either the Schwarzschild-coordinate-based derivation\cite{Kenyon1990,DInverno1992} or the one based on isotropic coordinates,\cite{Rindler2001} but not the discrepancy between the two. Historically, Shapiro's first derivation\cite{Shapiro1964} is based on Schwarzschild coordinates, while the formulae he later used to fit his observational results are those based on isotropic coordinates.\cite{Shapiro1971} The differences between the two have been discussed in depth in Ref.~\onlinecite{Ichinose1989}.

Returning to the simpler teaching strategy of deriving the Newtonian approximation, and analysing the additional information from the Schwarzschild metric, both the metric derivations show us directly where the factor 2 is coming from: By construction, our Newtonian derivation only included the relation between proper time $t$ and coordinate time $\tau$, which in both forms of the metric is encoded in the $\Dd t^2$ term. In our metric calculations, this term only contributes half of the final result for the term that later on becomes the pre-factor of the logarithmic delay term. The other half comes from the spatial part of the metric --- which our Newtonian derivation using the EEP could not capture, given its assumption that space is Euclidean. In short, half of the general-relativistic result comes from the time part, half from the space part. Newtonian gravity can be interpreted as a distortion of time only,\cite{Gould2016} and cannot capture the contributions from the distortion of space.

This shortcoming of the Newtonian version is a general feature of derivations based on the equivalence principle,\cite{Marsh1975,Gruber1988} and in complete analogy with Newtonian derivations of the deflection of light, where the neglect of non-Euclidean spatial geometry leads to the same factor 2 discrepancy.\cite{Lerner1997,Strandberg1986,Ferraro2003} 

\section{Applications: Solar System and Pulsars}
\label{Sec:Data}
Formulas (\ref{Eq:TimeDelay1}) and (\ref{Eq:TimeDelay2}) for one-way travel, corrected by the multiplication of the delay term with an overall factor 2 to go from the Newtonian to the general-relativistic result,
\be
\Delta t= \frac{2GM}{c_0^3}\ln\left[
\frac{x_E+r_E}{x_P+r_P}
\right] =\frac{2GM}{c_0^3}\ln\left[
\frac{(r_E +\vec{n}\cdot\vec{x}_E)(r_P -\vec{n}\cdot\vec{x}_P)}{\eta^2}\right],
\label{Eq:TimeDelayCorrected}
\ee
can readily be applied to specific situations in which the Shapiro time delay can be measured. 

Let us start with the typical situation for measurements within our solar system. In that case, the mass is the Sun, and we are looking at a situation where a planet is almost directly behind the Sun, as seen from Earth --- in other words: a situation close to what is known in astronomy as a superior conjunction of that planet, with the planet, the Sun and Earth all lined up, in that order. That is the situation where light travelling from us to the other planet or back will pass close to the Sun, and where the Shapiro time delay is particularly large.

\begin{figure}[htbp]
\begin{center}
\includegraphics[width=0.65\textwidth]{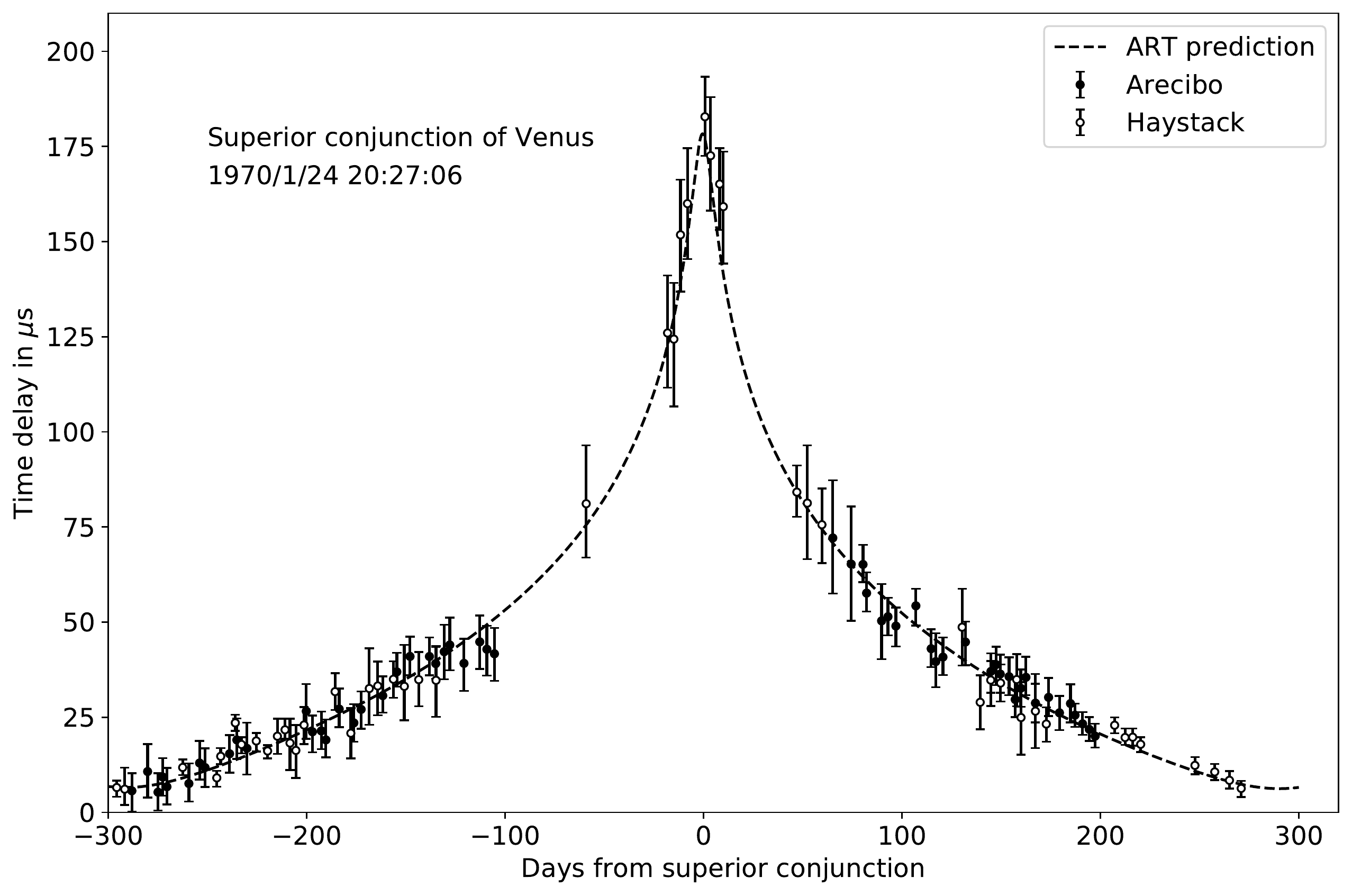}
\caption{Comparison between the prediction of (\ref{Eq:TimeDelayCorrected}) and Shapiro's radar echo data for Venus from 1970/1971}
\label{Fig:ShapiroData}
\end{center}
\end{figure}

Ephemeris software will typically give us the distances $r_E$ and $r_P$, plus the distance $r_{EP}$ between the planet and Earth, as well as the elongation angle $\alpha$, which is the angle $\angle SEP$ in Fig.~\ref{GTDSetup2}. From these quantities, we can derive $\eta=r_E\sin\alpha$, $x_E=r_E\cos\alpha$, and $x_P=x_E-r_{EP}$. Fig.~\ref{Fig:ShapiroData} shows the computed time delay for radar echoes sent to Venus before, during and after a particular superior conjunction. The light travel time is short compared to the rates of change in planetary positions, so the prediction for the Shapiro time delay for signals sent from Earth to Venus and reflected back to Earth is twice the delay predicted by (\ref{Eq:TimeDelayCorrected}), adding up the delay for the propagation from Earth to Venus and that from Venus to Earth for a single set of values for $r_P, r_E, r_{EP}$ and $\alpha$. The prediction is compared with selected data extracted from Fig.~1 in Shapiro's 1971 article, ref.~\onlinecite{Shapiro1971}. The ephemeris data was taken from the ephem Python package programmed by Brandon Rhodes.\cite{Rhodes}
The implementation is straightforward enough that it could form part of a programming-based astronomy lab course. The data I have extracted from the Shapiro article for this purpose is available online at Ref.~\onlinecite{OnlineShapiro}, together with the script I used to create Fig.~\ref{Fig:ShapiroData}.

A second interesting application is to binary pulsars, that is, to binary star systems where at least one of the stars is a pulsar. The first such system, the binary neutron star that contains the pulsar PSR B1913+16, was discovered by Joseph P. Taylor and Russell Hulse in 1974, and has enabled interesting tests of general relativity close to compact objects, including the first indirect detection of gravitational waves. 

\begin{figure}[htbp]
\begin{center}
\includegraphics[width=0.5\textwidth]{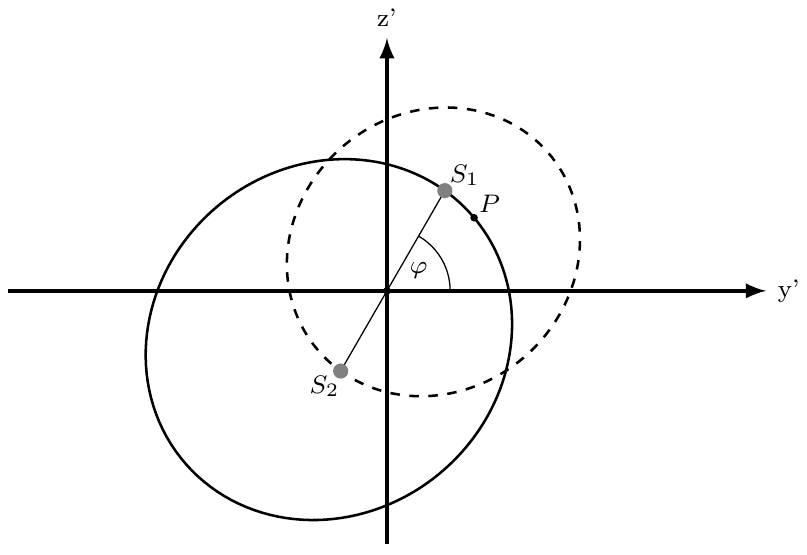}
\caption{Orbital plane of the binary containing a pulsar}
\label{PulsarSetupOrbit}
\end{center}
\end{figure}

The following derivation is a pedagogical elaboration of the time-delay formula first found by Blandford and Teukolsky.\cite{Blandford1976}
We begin by describing the orbital plane of the binary, as shown in Fig.~\ref{PulsarSetupOrbit}. 
The origin of the plane is the center of mass of the system, around which the two stars $S_1$ and $S_2$ revolve. Let the masses of the stars be $m_1$ and $m_2$, respectively. As shown in the diagram, the two stars are always on opposite sides of the origin. We describe the phase of their orbit by the angle $\varphi$. The value of the angle for when the first star is at the periapsis point $P$, that is, in the position where the mutual distance between the two stars is minimal, is called the {\em argument} (alternatively: {\em longitude}) {\em of pericenter}, commonly denoted by $\omega$.

The mutual distance between the stars can be written as
\be
r(\varphi) = \frac{a(1-e^2)}{1+e\cos(\varphi-\omega)},
\ee
using the standard polar coordinate formula for ellipses, where $a$ is the major half-axis and $e$ the numerical eccentricity of the ellipse that solves the reduced Kepler problem. The difference $\nu\equiv \varphi-\omega$ is called the {\em true anomaly}. Solving the two-body Kepler problem is a standard exercise in celestial mechanics. By Kepler's third law, $a$ can be expressed in terms of the orbital period $T$ and the system's total mass as
\be
a^3 = T^2\cdot\frac{G(m_1+m_2)}{4\pi^2},
\ee
while there is a multi-step process to obtain the $\varphi$ as a function of time: The true anomaly $\nu$ is linked to the so-called {\em eccentric anomaly} $E$ by
\be
\tan\nu = \frac{\sqrt{1 - e^2\,} \sin{E}}{\cos{E} -e},
\ee
while $E$ can be obtained from the {\em mean anomaly} $M$ by Kepler's equation
\be
M = E-e\sin E
\ee
(which cannot be solved analytically), and the mean anomaly can be written as a function of time as
\be
M =\frac{2\pi}{T}(t-t_0),
\ee
with $t_0$ the time when both stars are in the pericenter position. Since our origin is at the center of mass, the distances of the two stars from the origin are given by
\be
r_1(\varphi)=\frac{m_1}{m_1+m_2}r(\varphi), \;\;\;\; r_2(\varphi)=\frac{m_2}{m_1+m_2}r(\varphi),
\ee
and as short-hand, we can introduce the parameters $p_1$ and $p_2$ \be
p_{1,2} \equiv \frac{m_{1,2}}{m_1+m_2}\cdot a(1-e^2).
\ee
(The technical term for $p_{1,2}$ is that each is the ``semi-latus rectum'' of the elliptical orbit of the star in question.)

In the right-handed coordinate system in which the orbital plane of our stars is the y'-z' plane shown in Fig.~\ref{PulsarSetupOrbit}, the locations of the two stars as a function of $\varphi$ are
\be
\vec{x}{\,}'_1(\varphi) = \frac{p_1}{1+e\cos(\varphi-\omega)}\left(
\begin{array}{c}
0\\
\cos\varphi\\
\sin\varphi
\end{array}
\right),\;\;\;\;\; \vec{x}{\,}'_2(\varphi) = -\frac{p_2}{p_1}\cdot\vec{x}{\,}'_1(\varphi).
\ee
In summary, in order to describe the motion of the stars in the orbital plane, we need to know the epoch $t_0$, the total mass $m_1+m_2$, the eccentricity $e$, and the argument of the pericenter $\omega$. 

\begin{figure}[htbp]
\begin{center}
\includegraphics[width=0.8\textwidth]{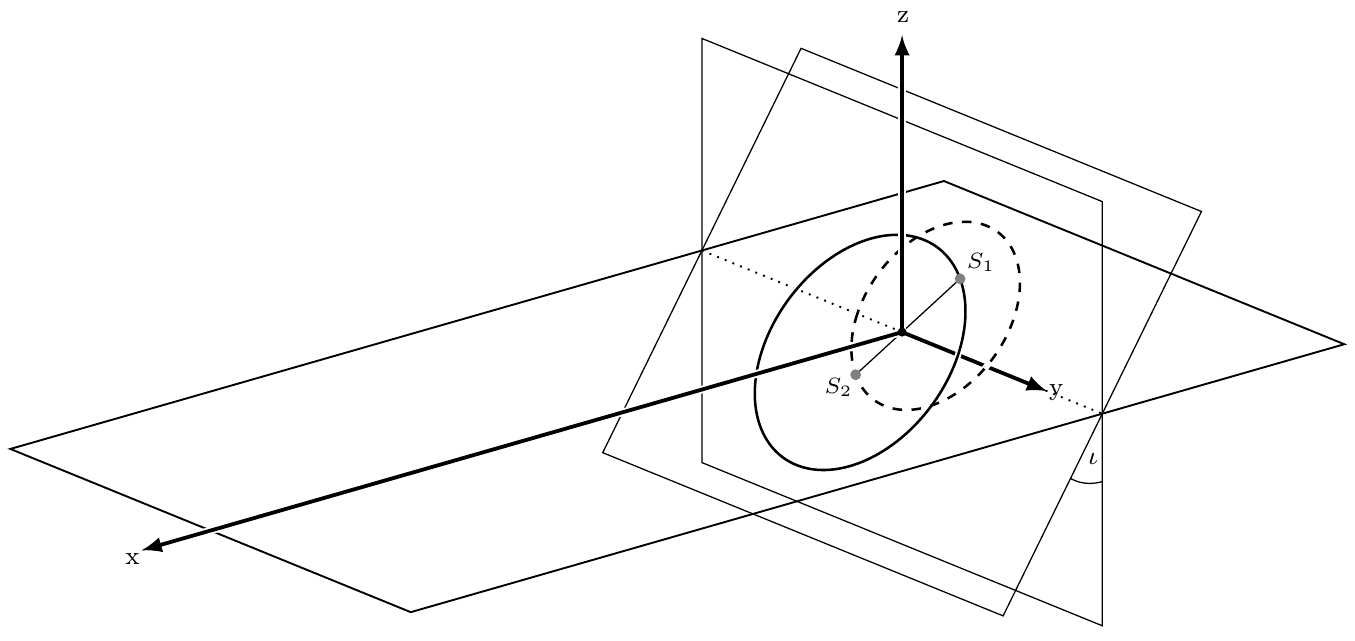}
\caption{Binary containing a pulsar, as seen from Earth. The orbital plane of the stars is inclined relative to the yz plane by the inclination angle $\iota$}
\label{PulsarSetup}
\end{center}
\end{figure}

As we observe the double star from afar, there is an additional parameter related to the orientation of the orbital plane relative to us. The three-dimensional set-up can be seen in Fig.~\ref{PulsarSetup}. We choose the center of mass of the system as the origin of our coordinate system, and the x axis as our line of sight from Earth, which we take to be located at $\vec{x}_E = (r_{CE},0,0)^T.$ We choose the y-z plane as shown in the figure, which amounts to choosing the longitude of the ascending node to be $\pi/2$. With this choice, we have one additional orbital parameter, namely the inclination $\iota$ of the orbital plane relative to the yz plane, where  $\iota=\pi/2$ corresponds to an edge-on and $\iota=0$ to a face-on view of the system. We rotate our y'-z' system within its plane so that the y' axis coincides with the y axis.
The parameters we have introduced so far are collectively known as the {\em Keplerian orbital elements}: time of epoch $t_0$, inclination $\iota,$ longitude of the ascending node (which we have chosen as $\pi/2$), orbital eccentricity $e$, semimajor axis $a$ and argument of longitude $\omega$. They are commonly used to specify orbits in classical mechanics. 

\begin{figure}[htbp]
\begin{center}
\includegraphics[width=0.5\textwidth]{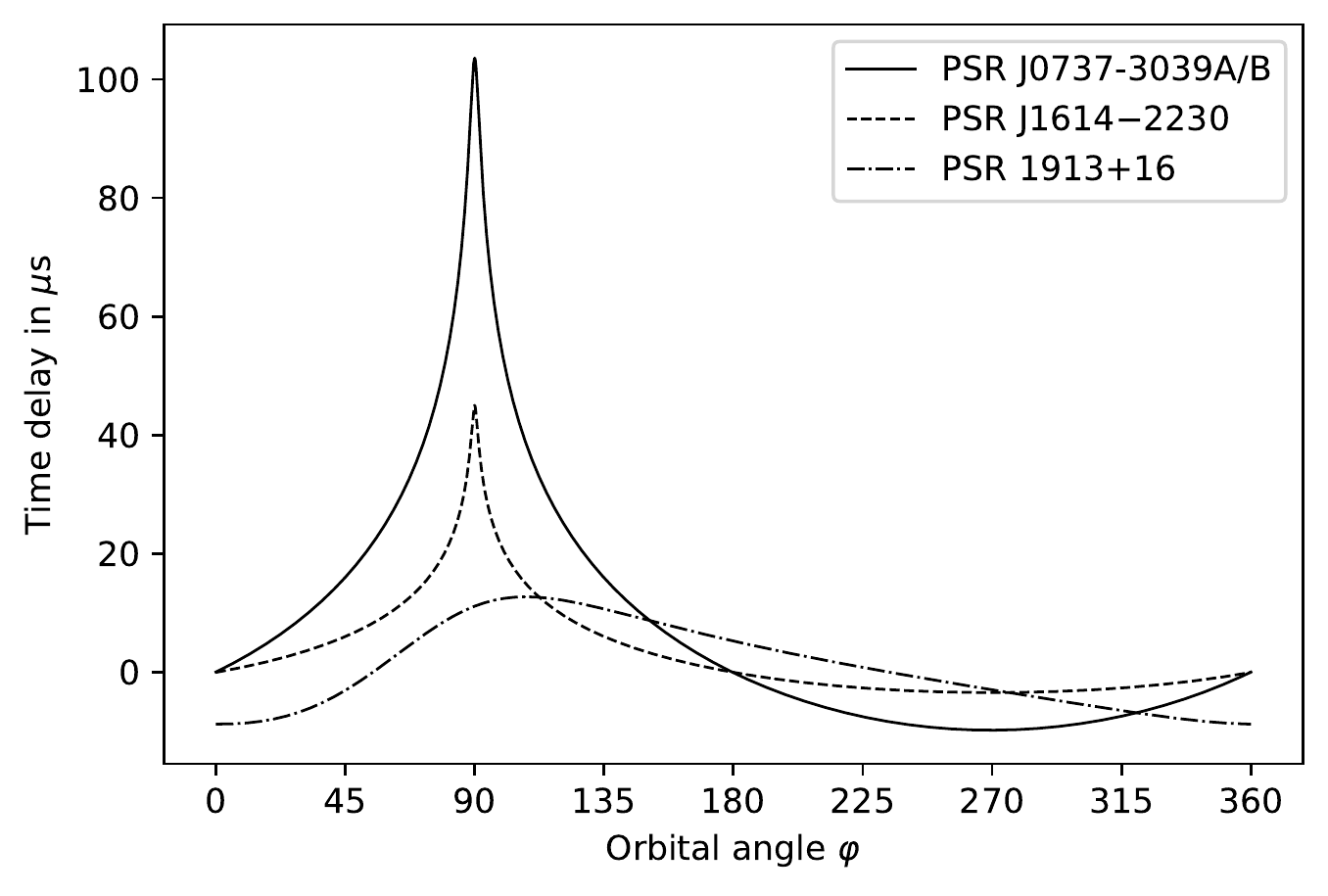}
\caption{Time-variable portion of the Shapiro time delay for PSR 1913+16, PSR J0737-3039A/B and PSR J1614-2230,
plotted by inserting best-fit parameters into (\ref{Eq:ShapiroDelayPulsars})}
\label{ShapiroPulsarData}
\end{center}
\end{figure}

From our definitions, it follows that the locations of stars 1 and 2 in our coordinate system are given by
\be
\vec{x}_1(\varphi) = \left(
\begin{array}{ccc}
\cos\iota & 0 & -\sin\iota\\
0 & 1 & 0\\
\sin\iota & 0 & \cos\iota
\end{array}
\right)\cdot \vec{x}{\,}'_1(\varphi) = 
\frac{p_1}{1+e\cos(\varphi-\omega)}\left(
\begin{array}{c}
-\sin\varphi\:\sin\iota\\
\cos\varphi\\
\sin\varphi\:\cos\iota
\end{array}
\right)\ee
and
\be\vec{x}_2(\varphi) = -\frac{p_2}{p_1}\:\vec{x}_1(\varphi).
\ee
Let star 1 be the pulsar whose regular signals we can observe, and star 2 the compact mass causing the Shapiro delay. In line with our previous conventions, the location of star 2 will be the origin of our coordinate system. We will not aim at a precise description, but will freely make suitable approximations. Notably, the length scales $p_{1,2}$ are both much smaller than the distance between the center of mass and Earth $r_{CE}$, so we will keep only the leading-order terms in $p_{1,2}/r_{CE}\ll 1$. In order to be able to apply formula (\ref{Eq:TimeDelayCorrected}) for the Shapiro time delay in this system, we need to shift the origin of our coordinates to the location of the second star. In that new system, the position vector of the Earth is approximately
\be
\vec{x}{\,}''_E(\varphi) = \vec{x}_E -\vec{x}_2(\varphi) = \left(
\begin{array}{c}
r_{CE} \\
0\\
0\end{array}
\right),
\ee
where we have only kept the leading (zeroth order) term in $p_2/r_{CE}\ll 1$. The position of the pulsar is
\be
\vec{x}{\,}''_1 =  \vec{x}_1(\varphi) -\vec{x}_2(\varphi) = \frac{a(1-e^2)}{1+e\cos(\varphi-\omega)}
\left(
\begin{array}{c}
-\sin\varphi\:\sin\iota\\
\cos\varphi\\
\sin\varphi\:\cos\iota
\end{array}
\right).
\ee
The normal vector pointing from the pulsar to Earth is approximately $\vec{n} = (1,0,0)^T$, again keeping only the leading (zeroth order) term in $p_1/r_{CE}\ll 1$. From the two position vectors and the normal vector, it follows that the argument in the left-hand version of the Shapiro time delay term in equation (\ref{Eq:TimeDelayCorrected}) is
\be
\frac{|\vec{x}{\,}''_E| + \vec{n}\cdot \vec{x}{\,}''_E}{|\vec{x}{\,}''_1|+\vec{n}\cdot \vec{x}{\,}''_1} = \frac{2r_{CE}}{a(1-e^2)}\cdot \frac{1+e\cos(\varphi-\omega)}{1-\sin\varphi\:\sin\iota}.
\label{Eq:PulsarFactors}
\ee
In the logarithm, the constant factor $2r_{CE}/a(1-e^2)$ produces a constant time shift that is undetectable, as we can only measure how the pattern of pulsar pulses varies over time, and not their absolute timing. Inserting the time-dependent factors of (\ref{Eq:PulsarFactors}) into (\ref{Eq:TimeDelayCorrected}), we find the time-dependent portion of the Shapiro delay to be 
\be
\Delta t(\varphi)= \frac{2Gm_2}{c_0^3}\ln\left[
\frac{1+e\cos(\varphi-\omega)}{1-\sin\varphi\:\sin\iota}
\right] = 2\:r_{Sh}\ln\left[
\frac{1+e\cos(\varphi-\omega)}{1-s_{Sh}\cdot \sin\varphi}
\right],
\label{Eq:ShapiroDelayPulsars}
\ee
where the second formula introduces the two parameters characteristic for the Shapiro time delay, namely the {\em range}
\be
r_{Sh} \equiv \frac{Gm_2}{c_0^3} \approx 5\:\si{\micro\second}\cdot\frac{m_2}{M_{\odot}}
\ee
that sets the basic time scale and the {\em shape}
\be
s_{Sh} \equiv\sin\iota.
\ee
The Shapiro range and shape parameters $r_{Sh}$ and $s_{Sh}$ are two of the so-called {\em post-Keplerian} parameters one can measure for general-relativistic orbits. The Shapiro contribution to the timing sequence is particularly pronounced both in the double pulsar PSR J0737-3039A/B, with two neutron stars that are both detectable as pulsars,\cite{Kramer2006} and the pulsar PSR J1614-2230 which orbits a White Dwarf star.\cite{Demorest2010} For PSR 1913+16, on the other hand, the marked deviation from the shape parameter from 1 smoothes out the effect, although $r_{Sh}$ and $s_{Sh}$ are fitted in at least some of the timing models. 
The shape parameters for those three cases, plus the Keplerian parameters necessary for calculating the Shapiro time delay, are listed in table \ref{BinaryNeutronTable}. 
\begin{table}
\bgroup
\renewcommand{\arraystretch}{1.2}
\begin{tabular}{c|c|c|c}
 & PSR 1913+16& PSR J0737-3039A/B & PSR J1614-2230\\\hline\hline
 $T$ [d] & $\;\;0.322997\;\;$ & $\;\;0.10225156248(5)\;\;$ & $\;\;8.6866194196(2)\;\;$\\\hline
 $e$ & $\;\;0.6171\;\;$ & $\;\;0.0877775(9)\;\;$ & $\;\;1.30(4)\cdot 10^{-6}\;\;$ \\\hline
 $\omega$ [deg] &  $220.14$ & $87.0331(8)$     &  $175(2)$  \\\hline
 $r_{Sh} [\si{\micro\second}]$ & $\;\;6.83\;\;$ & $\;\;6.21(33)\;\;$ & $\;\;2.46\;\;$ \\\hline
 $s_{Sh}$ & $\;\;0.734\;\;$ & $\;\;0.99974(-39,+16)\;\;$ & $\;\;0.999894(5)\;\;$
\end{tabular}
\egroup
\caption{Selected post-Keplerian and Keplerian orbital parameters for binary star systems containing at least one pulsar. PSR 1913+16 data from Tables and 5 [DD(1) and DD(3)] from ref.~\onlinecite{Taylor1989}; PSR J0737-3039A/B data from Table 2 in ref.~\onlinecite{Kramer2006}, PSR J1614-2230 derived from data from Table 1 in ref.~\onlinecite{Demorest2010}. The meaning of the parameters is explained in the text}
\label{BinaryNeutronTable}
\end{table}
The resulting time-variable Shapiro time delay as a function of the orbital angle $\varphi$ is plotted in Fig.~\ref{ShapiroPulsarData}. Given the small eccentricities for both PSR J0737-3039A/B and PSR J1614-2230, the same plot also gives a reliable impression of how the time delay changes over time in those two systems.

\section{Discussion}

We have shown how to derive the Shapiro time delay near a mass from the equivalence principle, at the same level of simplification as that employed in some derivations of the gravitational deflection of light. The key elements of the derivation were two sets of coordinates: one adapted to Newtonian physics (and with a non-constant speed of light), the other local free-fall coordinates in which the equivalence principle could be applied. As in the corresponding Newtonian derivations of the deflection angle of light, the overall result is too small by a factor of 2, indicating that while time distortion effects have been taken into account, the curvature of space has been neglected. Nevertheless, the result has the same functional form, with its characteristic logarithm, as more precise calculation, and as such can serve to introduce the gravitational time delay in an undergraduate or even high school setting.

While a derivation of the Shapiro time delay from the Schwarzschild metric readily showed the origins of the missing factor 2, it also demonstrated ambiguities in metric-based text book derivations of the Shapiro time delay (that are also present in the original research articles about the effect). Arguably, those ambiguities should be mentioned explicitly at the introductory text book level when a metric-based derivation of the gravitational time delay is presented, even if a rigorous treatment (that is, the proper post-Newtonian formalism) is beyond the scope of such texts. 

The time delay formula itself can readily be applied to observational data. In the solar system, fitting such data can be achieved by retrieving ephemeris data from suitable software. For solar system measurements involving Venus, the article includes a link to reconstructed observational data from Shapiro et al. 1970/1971. Fitting the data for a simple test of general relativity would be a suitable exercise for an undergraduate astronomy lab. For binary stars, the conventions for Keplerian and post-Keplerian orbital elements are introduced, enabling students to check for themselves some of the basic equations that are at the foundation of today's most precise tests of general relativity involving compact binary stars.

\section*{Acknowledgements}

I would like to thank Gerhard Sch\"afer for helpful discussions of the post-Newtonian formalism, and Thomas M\"uller for his valuable comments on an earlier version of this article.


\end{document}